\documentclass[
aps,
prx,
amsmath,amssymb,
superscriptaddress,
reprint,
nofootinbib,
floatfix,
longbibliography,
noeprint
]{revtex4-2}

\usepackage[english]{babel}
\usepackage[utf8]{inputenc}
\usepackage[T1]{fontenc}
\usepackage{physics}
\usepackage{amsmath}
\usepackage{mathrsfs}
\usepackage{bbm}
\usepackage{bm}
\usepackage{enumerate}
\usepackage{graphicx}
\usepackage[dvipsnames]{xcolor}
\usepackage{float}
\usepackage{subdepth}
\usepackage{makecell}
\usepackage{fnpct}
\usepackage{booktabs}
\usepackage{multirow}
\usepackage{setspace}
\usepackage[math]{cellspace}
\usepackage{subfigure}
\usepackage[normalem]{ulem}

\usepackage[colorlinks,
linkcolor=BrickRed,
citecolor=MidnightBlue,
urlcolor=MidnightBlue,
bookmarks=true,
bookmarksopen=true,
bookmarksnumbered=true,
]{hyperref}

\newcommand{\eref}[1]{(\ref{#1})}
\newcommand{\dob}{\begin{tabular}{c}}
\newcommand{\edob}{\end{tabular} }

\newcommand{\sign}{\mathrm{sign}}
 \newcommand{\wtilde}{\widetilde}
 \renewcommand{\P}{\mathbb{P}}

\renewcommand*\d{\mathop{}\!\mathrm{d}}

\renewcommand{\i}{\mathrm{i}}
\renewcommand{\(}{\left(}
\renewcommand{\)}{\right)}
\newcommand{\id}{\mathbbm{1}}

\newcommand{\sL}{\mathcal{L}}

\newcommand{\be}{\begin{eqnarray}}
\newcommand{\bea}{\begin{eqnarray}}
\newcommand{\eea}{\end{eqnarray}}
\newcommand{\beq}{\begin{equation}}
\newcommand{\ee}{\end{eqnarray}}
\newcommand{\eeq}{\end{equation}}

 \newcounter{ls}

\newcounter{amg}

\newcounter{jvc}

\newcounter{yc}

\begin{document}
\title{Emergent topology in many-body dissipative quantum matter}

\author{Antonio M. Garc\'\i a-Garc\'\i a}
\email{amgg@sjtu.edu.cn}
\affiliation{Shanghai Center for Complex Physics,
	School of Physics and Astronomy, Shanghai Jiao Tong
	University, Shanghai 200240, China}

\author{Lucas S\'a}
\email{ld710@cam.ac.uk}
\affiliation{TCM Group, Cavendish Laboratory, University of Cambridge, JJ Thomson Avenue, Cambridge CB3 0HE, UK\looseness=-1}
\affiliation{CeFEMA, Instituto Superior T\'ecnico, Universidade de Lisboa, Av.\ Rovisco Pais, 1049-001 Lisboa, Portugal}

\author{Jacobus J. M. Verbaarschot}
\email{jacobus.verbaarschot@stonybrook.edu}

\affiliation{Center for Nuclear Theory and Department of Physics and Astronomy, Stony Brook University, Stony Brook, New York 11794, USA}
\author{Can Yin}
\email{yin_can@sjtu.edu.cn}
\affiliation{Shanghai Center for Complex Physics,
	School of Physics and Astronomy, Shanghai Jiao Tong
	University, Shanghai 200240, China}

\date{\today}

\begin{abstract}
The identification, description, and classification of topological features is an engine of discovery and innovation in several fields of physics.
This research encompasses a broad variety of systems, from the integer and fractional Chern insulators in condensed matter, to protected states in complex photonic lattices in optics, and the structure of the QCD vacuum.
Here, we introduce another playground for topology: the dissipative dynamics of pseudo-Hermitian many-body quantum systems. For that purpose, we study two different systems,
the dissipative Sachdev-Ye-Kitaev (SYK) model,
and a quantum chaotic dephasing spin chain.
For the two different many-body models, we find the same topological features for a wide
range of parameters suggesting that they are universal.
In the SYK model,
we identify four universality classes, related to pseudo-Hermiticity, characterized by a rectangular block representation of the vectorized Liouvillian that is directly related to the existence of an anomalous trace of the unitary operator implementing fermionic exchange. As a consequence of this rectangularization, we identify a topological index $\nu$ that only depends on symmetry. Another distinct consequence of the rectangularization is the observation, for any coupling to the bath, of purely real topological modes in the Liouvillian.
The level statistics of these real modes agree with that of the corresponding random matrix ensemble and therefore can be employed to characterize the four topological symmetry classes.
In the limit of weak coupling to the bath, topological modes govern the approach to equilibrium, which may enable a direct path for experimental confirmation of topology in dissipative many-body quantum chaotic systems.
\end{abstract}

\maketitle

\section{Introduction}

Nontrivial topological features in wave functions and gauge configurations
have helped explain a broad variety of physical phenomena in different
research areas. The anomalous decay of the neutral pion into two photons due to the chiral anomaly~\cite{adler1969,jackiw1969} has its origin~\cite{thooft1976,nielsen1977} in certain topologically nontrivial gauge configurations of zero field strength in which fermions have zero modes.
The Atiyah-Singer index theorem~\cite{atiyah1968,alvarez1983supersymmetry} then relates the number of the fermionic zero modes with a topological invariant of the gauge fields. Later, these ideas played an important role in the development of effective models of the QCD vacuum aimed at explaining the phenomenon of spontaneous chiral symmetry breaking~\cite{shuryak1982}, and also in the proposal of the chiral random matrix ensembles~\cite{verbaarschot1993a,verbaarschot1994} that capture universal properties of the QCD Dirac operator. In condensed matter, the experimental discovery~\cite{klitzing1980} of the quantization of the Hall conductivity and its theoretical explanation~\cite{thouless1982} based on the existence of topologically protected edge states triggered an enormous interest in topological quantum matter. Highlights of this endeavor include the discovery~\cite{tsui1982} and theoretical explanation~\cite{laughlin1983} of the fractional quantum Hall effect caused by strong electronic interactions in two dimensions and the theoretical prediction~\cite{haldane1988}, and experimental confirmation~\cite{konig2007}, of topological features without the need of a magnetic field, the so-called spin-Hall effect~\cite{kane2005,kane2005a,bernevig2006}.

The realization that, in momentum space, topology only requires a certain symmetry in the band structure of the material, which depends strongly on the dimensionality, and the existence of a gap in the spectrum that protects the edge states led to the proposal~\cite{fu2007,fu2007a,roy2009,moore2007}, and experimental confirmation~\cite{dqian2008}, of topological insulators in three dimensions.
Such ideas were also applied to the characterization of topological superconductors~\cite{sato2017}, and the employment of K-homology techniques led to a full characterization of topological insulators and superconductors~\cite{kitaev2009,ryu2010,schnyder2008} based on symmetry and dimensionality.
In this context, interactions can, in some cases~\cite{kitaev2010}, change the nature of the topological invariants~\cite{kitaev2010}.

Topology has also been intensively investigated in non-Hermitian Hamiltonians or Liouvillians in condensed matter and cold atoms~\cite{esaki2011,rudner2009,hu2011,okuma2023,gong2018,ueda2019,shen2018,yao2018,schomerus2013,kunst2018,slager2020,okuma2020,schindler2023}, in the context of QCD Wilson fermions
\cite{Itoh:1987iy, Golterman:1992ub,Narayanan:1994gw,Kaplan:2009yg,Akemann:2010em},
and in random matrix theory~\cite{ Osborn:2004rf,Akemann:2004dr,Kanazawa:2009en,Damgaard:2010cz,Akemann:2010em,Kieburg:2011uf,Akemann:2011kj,Kieburg:2013xta,Kieburg:2015vqa}, while quantum optical settings~\cite{schomerus2013} are an adequate experimental platform to simulate non-Hermitian topology.
In condensed matter and cold atoms, most of the research~\cite{gong2018,esaki2011} has focused on noninteracting systems with an emphasis on comparing with the previously mentioned classification of noninteracting topological insulators by using, for instance, Hermitization techniques.
Preliminary studies of the role of interactions and topology
in non-Hermitian systems~\cite{ryu2022,zhang2022}, employing the interacting Hatano-Nelson model~\cite{hatano1996,faugno2022} and Liouvillians with two-body losses~\cite{hamanaka2023}, point to profound differences with respect to the noninteracting limit~\cite{yoshida2023} that call for new ideas and techniques.

In this paper, we propose a novel platform to study non-Hermitian topology in strongly interacting systems: open many-body quantum systems with pseudo-Hermitian symmetry, which are characterized by a topological index that only depends on symmetry. They belong to four universality classes that are topological extensions of four universality classes in the Bernard-LeClair~\cite{bernard2002} classification
of non-Hermitian random matrices. To keep the discussion concrete, we focus mostly on the dissipative dynamics of the Sachdev-Ye-Kitaev (SYK) model~\cite{bohigas1971,french1970,bohigas1971a,french1971,mon1975,kitaev2015,maldacena2016}
coupled to a Markovian bath~\cite{sa2022,kulkarni2022,kawabata2022,garcia2022e,garciapengfei}, but to illustrate the universality of the topological features we also
study a dephasing spin chain with nearest-neighbor interactions~\cite{sa2022a}.

The remainder of the paper is organized as follows. We start our analysis in Sec.~\ref{sec:model} with the definition of the model and the description of its symmetries, in particular the pseudo-Hermiticity from which many-body topology emerges. While the discussion is particular to the SYK model, every important concept can be readily extended to more general settings. We then introduce a topological index, which is computed analytically (Sec.~\ref{sec:analytical}) and numerically (Sec.~\ref{sec:numerical}), and relate it to the existence of purely real decay modes of the Lindbladian. Sec.~\ref{sec:universality} is devoted to the universality of topology in several directions, in particular in non-SYK models, the spectral statistics of real modes, and experimentally relevant dynamical quantities.
We conclude in Sec.~\ref{sec:conclusion} by summarizing our findings and propose further directions of research.

\section{Model and symmetries}
\label{sec:model}

We study the SYK model of $N$ Majoranas $\psi_i=\psi_i^\dagger$, $i=1,2, \dots, N$, with $q$-body interactions of infinite range in zero dimensions, whose Hamiltonian is given by
\begin{equation}
	H = -i^{q/2}\sum_{1\leq i_1<i_2\cdots <i_q\leq N} K_{i_1 i_2\cdots i_q}\psi_{i_1}\psi_{i_2}\cdots \psi_{i_q },
	\label{eq:single_syk}
\end{equation}
where $\{\psi_i,\psi_j\} =\delta_{ij}$, $K_{i_1 i_2\cdots i_q}$ are Gaussian random couplings
 with zero average and variance $J^2 2^{q-1}(q-1)!/(q N^{q-1})$, $N$ is even, and $q$ is a multiple of four. Throughout, we set the interaction strength to $J=1$. We are mainly interested in the dynamics of this SYK coupled to a Markovian bath by $r$-body jump operators $L_{i_1 i_2 \cdots i_r}=i^{r(r-1)/2}\psi_{i_1} \psi_{i_2}\cdots \psi_{i_r}$, described by the time evolution of the density matrix $\rho$,
\begin{align}
  \frac{d\rho}{dt}&  = \mathcal{L}(\rho) \equiv -i\left[H, \rho\right]
  + \lambda \frac{N^{1-r}}{r}\\
  \times&\sum_{1 \leq i_1<\cdots<i_r\leq N}\!\left(L_{i_1 \cdots i_r} \rho L_{i_1 \cdots i_r}^{\dagger}-\frac{1}{2}\{L_{i_1 \cdots i_r}^{\dagger} L_{i_1 \cdots i_r}, \rho\}\right)\!, \nonumber
  \end{align}
where $\cal L$ stands for the Liouvillian (of the Lindblad form) and $\lambda$ denotes the dissipation strength.
After the necessary vectorization to set up the path integral on the Keldysh contour~\cite{sa2022,kulkarni2022,garcia2022e}, with branches denoted by $+$ and $-$, namely, $\psi_k \rho \to \psi_k^+ \ket{\rho}$ and $\rho\psi_k\to -i\psi_k^-\ket{\rho}$, where $|\rho\rangle=\sum_{ij}\langle i| \rho |j\rangle |i\rangle |j\rangle$ is the vectorized density matrix, the vectorized Liouvillian is given by,
\begin{equation}
	  \mathcal{L}
          = -i H_+ + iH_- +\lambda H_{I},
	\label{eq:Keldysh_twosite SYK}
\end{equation}
where $H_{+,-}$ are obtained from the single-site SYK Hamiltonian in Eq.~(\ref{eq:single_syk}) by replacing $\psi_k\to \psi_k^{+,-}$
with identical couplings for
$H_+ = H\otimes \mathrm{I}$ and $H_- = {\mathrm I} \otimes H$, where $\rm{I}$ is the identity matrix, while
\begin{equation}
  H_{I} = i^r\frac{N^{1-r}}{r}\sum_{1\leq i_1<\cdots<i_r\leq N}
  \psi^+_{i_1}\cdots \psi^+_{i_r}\psi^-_{i_1}\cdots\psi^-_{i_r}
  \label{eq:hi}
\end{equation}
describes the effect of the bath.
We have dropped the constant term $-\lambda{N^{1-r} 2^{-r}}{N\choose r}/r$, which ensures the trace preservation of the density matrix (conservation of probability), in Eq.~(\ref{eq:hi}). This is advantageous for symmetry considerations, as it centers the spectrum at the origin.
For later use, we define $H_0=-iH_+ +iH_-$.
We note that the Liouvillian above has been recently employed~\cite{kawabata2022a} in a symmetry classification of many-body quantum chaotic systems in contact with a Markovian bath.

Any Liouvillian satisfies the reality condition $Q\sL^*=\sL Q$, which reflects the preservation of the Hermiticity of $\rho$ under the dissipative dynamics~\cite{sa2022a} (also known as PT symmetry in some contexts~\cite{garcia2023c}), and where $Q$ is a unitary operator that exchanges the $+$ and $-$ contours. More precisely, for the SYK model,
\begin{equation}
	\label{eq:def_Q}
	Q=\exp\left\{-\frac{\pi}{4}\sum_{i=1}^N\psi_i^+\psi_i^-\right\}
	=\prod_{i=1}^N\frac{1}{\sqrt{2}}\(1-2\psi_i^+\psi_i^-\)
\end{equation}
exchanges Majoranas on the $+$ and $-$ contours, up to a sign:
$Q\psi_i^+Q^{-1}=\psi_i^-$ and $Q\psi_i^- Q^{-1}=-\psi_i^+$. $Q$ anticommutes (commutes) with $H_0$ ($H_I$) and satisfies
$Q^4 = {\mathrm I}$, i.e., has eigenvalues $\pm1$ and $\pm i$.
For reasons that will become apparent shortly, our interest is focused
on values of $N$, $q$, and $r$ in Eqs.~(\ref{eq:Keldysh_twosite SYK}) and (\ref{eq:hi}) that lead to a pseudo-Hermitian Liouvillian under the same operator $Q$: $Q {\cal L}^\dagger = {\cal L} Q$. In a recent classification of PT-symmetric systems~\cite{garcia2023c}, we found four of these pseudo-Hermitian classes: AIII$_\nu$, BDI$^{\dagger}_\nu$, CI$_{--\nu}$, and BDI$_{++\nu}$; see Table~\ref{tab:topology_class} for details.

\begin{table*}[t]
  \caption{Rectangular block representation of the SYK Liouvillian Eq.~(\ref{eq:Keldysh_twosite SYK}) for pseudo-Hermitian classes AIII$_\nu$, BDI$^\dag_\nu$, BDI$_{++\nu}$, and CI$_{--\nu}$. In the second column, we tabulate the block structure~\cite{garcia2023c}, where the diagonal blocks have different sizes, namely, A is $(n+\nu)\times(n+\nu)$ and D is $n\times n$, while the off-diagonal blocks are rectangular, namely, B is $(n+\nu)\times n$ and C is $n\times(n+\nu)$. The third column gives the RMT universality class of the level statistics of the real topological modes in the limit of small $\lambda$. We also list the parameters for which the dissipative SYK model realizes the four topological classes, namely, the parity of $N/2$ and $r$ and the quantum numbers of the blocks with nonzero index $\Tr\mathbb P Q = \nu$. We stress that in all cases $q/2$ is even as this is a necessary condition for topology in the dissipative SYK model.}
	\begin{tabular}{|c|c|c|c|c|c|c|}
		\hline
		\multirow{2}{*}{Class} & \multicolumn{2}{c|}{Matrix realization} & \multicolumn{4}{c|}{Dissipative SYK}\\
		\cline{2-7}
		 & Block structure & RMT & $(-1)^{N/2}$ & $(-1)^r$ & Block & $\nu$ \\
		\hline
		AIII$_\nu$ & $\left(\begin{array}{cc}
		\mathrm{A} & \mathrm{B} \\
		-\mathrm{B}^{\dagger} & \mathrm{D}
		\end{array}\right),\quad \makecell[l]{\mathrm{A}=\mathrm{A}^{\dagger}\\ \mathrm{D}=\mathrm{D}^{\dagger}}$ & GUE &$-1$ & $+1$ & $S_+ S_-=1$ &$2^{\frac{N}{2}-1} $
		\\
		\hline
		BDI$^\dag_\nu$ &$\left(\begin{array}{cc}
		\mathrm{A} & \mathrm{B} \\
		-\mathrm{B}^{\dagger} & \mathrm{D}^*
		\end{array}\right),\quad\makecell[l]{\mathrm{A}=\mathrm{A}^{\top}=\mathrm{A}^{\dagger}\\ \mathrm{B}=\mathrm{B}^*\\ \mathrm{D}=\mathrm{D}^{\top}=\mathrm{D}^{\dagger}}$
		& GOE & $+1$ & $+1$ & $S_+ S_-=1$ &$ 2^{\frac{N}{2}-1}$
		\\
		\hline
		BDI$_{++\nu}$ &$\left(\begin{array}{cccc} & & \mathrm{A} & \mathrm{B} \\ & & \mathrm{C} & \mathrm{D} \\ \mathrm{A}^{\top} & \mathrm{C}^{\top} & & \\ \mathrm{B}^{\top} & \mathrm{D}^{\top} & & \end{array}\right),\quad \makecell[l]{ \mathrm{A}=\mathrm{A}^* \\  \mathrm{B}=-\mathrm{B}^* \\ \mathrm{C}=-\mathrm{C}^* \\ \mathrm{D}=\mathrm{D}^*}$
		&BDI & $+1$ & $-1$ & $S=1$& $2^\frac{N}{2}$
		\\
		\hline
		CI$_{--\nu}$ &$\left(\begin{array}{cccc}
		0 & 0 & \mathrm{~A} & \mathrm{~B} \\
		0 & 0 & -\mathrm{B}^{\top} & \mathrm{D} \\
		\mathrm{A}^* & \mathrm{~B}^* & 0 & 0 \\
		-\mathrm{B}^{\dagger} & \mathrm{D}^* & 0 & 0
		\end{array}\right)$,\quad \makecell[l]{$\mathrm{A}^{\top}=\mathrm{A}$\\ $\mathrm{D}^{\top}=\mathrm{D}$ }
		& CI & $-1$ & $-1$ & $S=1$&$2^{\frac{N}{2}}$
		\\\hline
	\end{tabular}
	\label{tab:topology_class}
\end{table*}

A key feature of these classes is the emergence of rectangular blocks of the SYK Liouvillian in sectors of good quantum numbers.
To see this, we define the single-site fermionic parities
\begin{equation}
\begin{split}
&S_+=2^{N/2} i^{N(N-1)/2}\prod_{i=1}^N\psi^+_i,\\
&S_-=2^{N/2} i^{N(N+1)/2}\prod_{i=1}^N i\psi^-_i,
\end{split}
\end{equation}
and the total fermionic parity $S = S_+ S_-$.
The Liouvillian block diagonalizes in a basis spanned by $S$ for odd $r$ and by $S_{+}$ and $S_-$ for even $r$ (that is, $\mathcal{L}$ [Eq.~\eref{eq:Keldysh_twosite SYK}] has a so-called strong symmetry~\cite{buca2012} $S_{+,-}$ when $r$ is even --- they both commute with $H_0$ and the dissipative term $\lambda H_I$ in the in the vectorized Lindbladian --- and a so-called weak symmetry~\cite{buca2012} $S$ when $r$ is odd --- in this case, $S_\pm$ no longer commutes with $H_I$).
Each of these blocks has additional subblock structure because of the commutation properties of $Q$ with $H_0$ and $H_I$ (we emphasize that $Q$ and $\mathcal{L}$ do not commute): in the basis in which $Q$ is diagonal, $H_I$ is block diagonal, while $H_0$ is block antidiagonal (i.e., $H_I$ connects states with the same eigenvalue of $Q$, while $H_0$ connects states with $Q=+1$ to states with $Q=-1$). In blocks with $S=+1$, corresponding to the sector of $Q$ with eigenvalues $\pm 1$, the trace of $Q$ (projected onto $S$ with $\P_S^s$ for odd $r$, or onto $S_+$ and $S_-$ for even $r$, using projectors $\mathbb{P}_{S_-}^{s_-}$ and $\mathbb{P}_{S_+}^{s_+}$), denoted by $\nu$, is nonzero: $\nu=\Tr  \P_S^{+1} Q = 2^{N/2}$ (odd $r$) and $\nu=\Tr \P_{S_-}^{\pm1} \P_{S_+}^{\pm1} Q = 2^{N/2}/2$ (even $r$).
This has the remarkable consequence that the two diagonal subblocks (corresponding to the two sectors of $H_I$ with $Q=\pm1$, respectively) have different dimensions, $n$ and $n+\nu$, with $2n+\nu=2^{N}/2$ for odd $r$ (i.e., $n=2^N-2^{N/2}/2$) and $2n+\nu=2^N/4$ for even $r$ (i.e., $n=2^N/8 -2^{N/2}/4$), and the off-diagonal subblocks (corresponding to $H_0$ and connecting states $Q=+1$ with $Q=-1$) become \emph{rectangular} $n\times (n+\nu)$ matrices, see Table~\ref{tab:topology_class}. At $\lambda=0$, each block of $\mathcal{L}$ is thus itself block antidiagonal with rectangular subblocks and has, therefore, $\nu$ zero modes. For finite $\lambda$, the pseudo-Hermiticity implies that the zero modes become purely real modes, as shown in Ref.~\cite{Kieburg:2013xta} for a matrix with the block structure of AIII$_\nu$ (see Table~\ref{tab:topology_class}).

The existence of rectangular blocks for the pseudo-Hermitian Liouvillian Eq.~(\ref{eq:Keldysh_twosite SYK}), closely related to the anomalous trace of the operator $Q$,
$\Tr \mathbb{P} Q = \nu$, suggests the existence of topological features. We now show explicitly that
the finite trace of $Q$ is indeed a topological invariant by an explicit calculation of the spectral flow of the eigenvalues of $Q({\cal L} +m \mathrm I )$ with ${\cal L}$ the vectorized
Liouvillian Eq.~(\ref{eq:Keldysh_twosite SYK}) and $m$ a real parameter that characterizes the flow.

\section{Analytic calculation of the topological index}
\label{sec:analytical}

We compute the index from the spectral flow of the $m$-deformed Liouvillian Eq.~(\ref{eq:Keldysh_twosite SYK}),
\begin{equation}
\wtilde{\cal L}(m)	=H_0+\lambda H_I+m \mathrm I, \label{eq:calli}
\end{equation}
where $m$ is real. [We consider $\wtilde{{\cal L}}(m)$ projected onto a sector with fixed quantum numbers, namely, $S=+1$ or $S_+=S_-$.]
$\wtilde{{\cal L}}$ is pseudo-Hermitian\footnote{Alternatively, $\mathcal{L}$ has the
noninvertible antiunitary involutive symmetry~\cite{komargodski2021symmetries,choi2023noninvertible}, $\mathbb{P}Q  \mathcal{L} =
\mathcal{L}^\dagger\mathbb{P}Q$.} (i.e., $Q\wtilde{\mathcal{L}}$ is Hermitian) and, hence, we consider the auxiliary Hermitian eigenvalue problem
\begin{equation}
	\label{eq:ekm}
	Q(H_0+\lambda H_I+m\mathrm{I})\ket{k}=E_k(m)\ket{k},
\end{equation}
or, equivalently,
	$(H_0+\lambda H_I)\ket{k}=-m\ket{k}+E_k(m)Q\ket{k}$,
where we used the fact that we are in the sector with $S=Q^2=+1$.
It follows that $-m$ is a real eigenvalue of the Lindbladian when $E_k(m)=0$ and, conversely, every real eigenvalue of $\mathcal{L}$ corresponds to an intersection of some flow line $E_k(m)$ with the $m$ axis.

The number of intersections of the flow lines with the $m$ axis, weighted by the sign of the slope at the intersection, is a topological
invariant $W$~\cite{Itoh:1987iy},
\begin{equation}
\begin{split}
W &= \sum_{k,\ell}\int_{-\infty}^{+\infty} \d m\ \delta(m-m_{k}^{(\ell)} )\ \sign \frac {\d E_k(m)}{\d m}
\\
&=\frac{1}{2}\sum_k \int_{-\infty}^{+\infty} \d m\ \frac {\d}{\d m}\sign(E_k(m)),
\label{tpinv}
\end{split}
\end{equation}
where $m_{k}^{(\ell)}$ is the $\ell$th intersection point of the flow line $E_k(m)$ with the $m$ axis. (Alternative integral representations for $W$ exist~\cite{ryu2022}.)
$W$ depends only on $E_k(m)$ evaluated at the endpoints $m \to \pm \infty$ and, in particular, is independent from $\lambda$. This follows immediately by evaluating the right-hand side of Eq.~(\ref{tpinv}) because, for $|m|\to\infty$, the eigenvalues of $Q \wtilde {\cal L}$ approach the eigenvalues of $mQ$. Since in the sector $Q^2=S=+1$, $Q$ has eigenvalues $\pm1$, we obtain
\be
W= \Tr \mathbb{P} Q  \equiv \nu(N,r),\label{eq:topind}
\ee
where $\mathbb{P}$ is the projector onto the relevant sector and the value of the topological index $\nu(N,r)$ is given in Table~\ref{tab:topology_class}.

It is instructive to compute $W$ explicitly for $\lambda=0$.
Since each intersection point of $E_k(m)$ with the real axis corresponds to a real eigenvalue of $Q \wtilde {\cal L}$, we can write
\be
W = \sum_{\{k| E_k=0 \}}  \sign \frac {\d E_k(m)}{\d m}.
\label{eq:def_W}
\ee
To proceed, we compute $\d E_k/\d m$ by applying $\bra{k}$ on the left of Eq.~(\ref{eq:ekm}), and taking a derivative with respect to $m$.
Then, we insert the corresponding result back into Eq.~(\ref{eq:def_W}), which yields
\begin{equation}\label{eq:def_W_sum}
W=\sum_{\{k| E_k=0 \}}\mathrm{sign}\bra{k}Q\ket{k}.
\end{equation}
Let $\ket{\zeta}$ be the eigenstates of $\sL=H_0$ with eigenvalue $\zeta$ at $\lambda=0$. We have
$\zeta\bra{\zeta}Q\ket{\zeta}=\bra{\zeta}\mathcal{L}Q\ket{\zeta}=-\bra{\zeta}Q\mathcal{L}\ket{\zeta}=-\zeta\bra{\zeta}Q\ket{\zeta}$,
so that $\bra{\zeta}Q\ket{\zeta}$ vanishes unless $\zeta=0$. When $\bra{\zeta}Q\ket{\zeta}$ does not vanish, we have that
\be
Q|\zeta\rangle = \alpha |\zeta\rangle  +|\delta \zeta\rangle,
\label{qeq}
\ee
where $\alpha$ is a constant and $ \langle \zeta | \delta \zeta\rangle = 0$. In general, there can be multiple zero modes, but we normalize
them such that $\langle k| Q|p\rangle \sim \delta_{kp}$ so that $ |\delta \zeta\rangle $ in Eq.~\eref{qeq} is perpendicular to all zero modes. Acting with $\bra{\zeta}$, we determine $\alpha$ and obtain
\be
Q|\zeta\rangle = \langle \zeta|Q |\zeta \rangle  |\zeta\rangle  +|\delta \zeta\rangle.
\ee
Acting with $H_0$ on both sides of this equation we find that $H_0|\delta \zeta\rangle = 0$, but
$|\delta \zeta\rangle$ is perpendicular to all zero modes. Therefore, $|\delta \zeta\rangle = 0$.
Acting a second time with $Q$, we find that
\be
 \langle \zeta|Q |\zeta \rangle ^2 =1.
\ee
At $\lambda=0$, we can, therefore, replace
$\mathrm{sign}\bra{k}Q\ket{k}$ in Eq.~(\ref{eq:def_W_sum}) by $\bra{k}Q\ket{k}$, and
extend the sum to all eigenstates $\ket{k}$ to arrive at Eq.~(\ref{eq:topind}).

The number of real eigenvalues
is at least equal to, and can be larger than, the difference of the number of flow lines from $-\infty$ to $+\infty$ and from $+\infty$ to $-\infty$ (the topological index $\nu$).
The number of real eigenvalues can change by a multiple of two when there are additional crossings. For example, if, upon varying $\lambda$, a flow line has $2n$ additional crossings, the number of real eigenvalues changes by $2n$, corresponding to $n$ complex-conjugated pairs of eigenvalues entering the real line; upon additional variation of $\lambda$, these pairs may leave the real line again.
Also, a true crossing at $\lambda=0$ can be deformed continuously into two lines with avoided crossings, resulting in two less real eigenvalues as $m$ goes from $-\infty$ to $+\infty$.
We have, therefore, two sources of real eigenvalues: $\nu$ topological ones coming from the rectangular structure of the Lindbladian, which are robust to changes in $\lambda$; plus those, sensitive to $\lambda$, that result from pairs of complex-conjugated eigenvalues entering and leaving the real line.

To summarize, we showed that the topological index $W$ is equal to the
index $\Tr \mathbb{P}Q$, which in our setting is closely related to the number of real modes in the small-$\lambda$ limit and, in turn, coincides with the parameter $\nu$ that controls the rectangularity of the block structure of ${\cal L}$.
This is our main finding.
In the next section, we present numerical results that illustrate our results, including the independence of $W$ from $\lambda$.
We stress that the existence of this topological invariant is an emergent property, due to an anomalous trace of the pseudo-Hermiticity operator, that cannot be easily guessed from the initial Liouvillian.
We also emphasize that the computation did not rely on any particular property of the SYK model besides a nonzero trace of $\mathbb{P}Q$. It is, therefore, valid for any other open many-body system with the same pseudo-Hermiticity symmetry under an appropriately defined $Q$.

\begin{figure*}[t]
	\centering
	\subfigure[]{\includegraphics[width=0.49\textwidth]{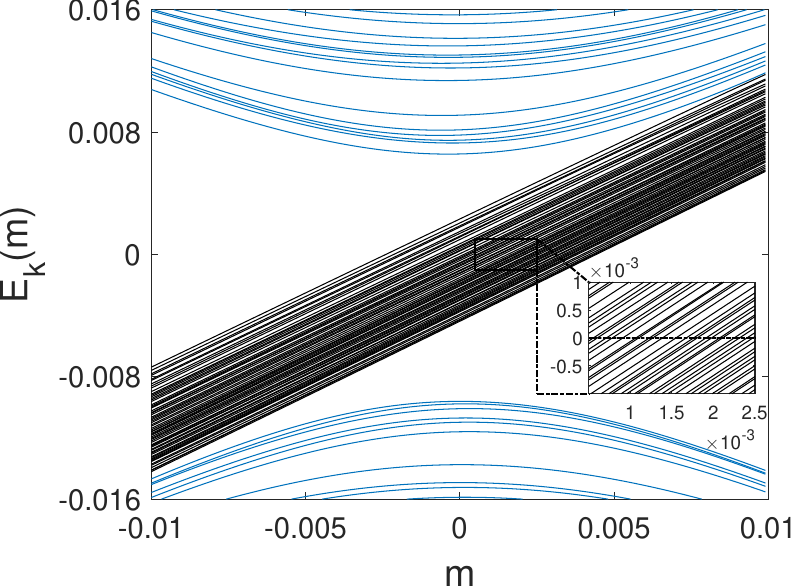}}
	\subfigure[]{\includegraphics[width=0.49\textwidth]{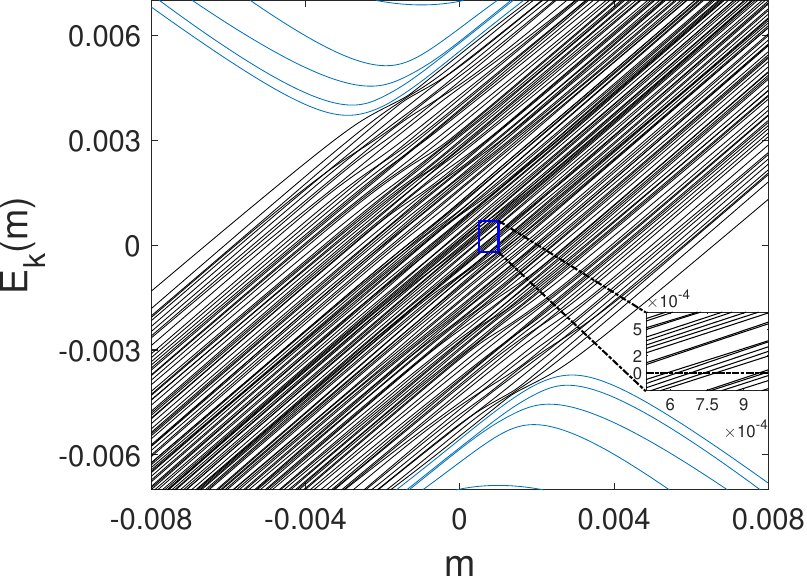}}\\
	\subfigure[]{\includegraphics[width=0.49\textwidth]{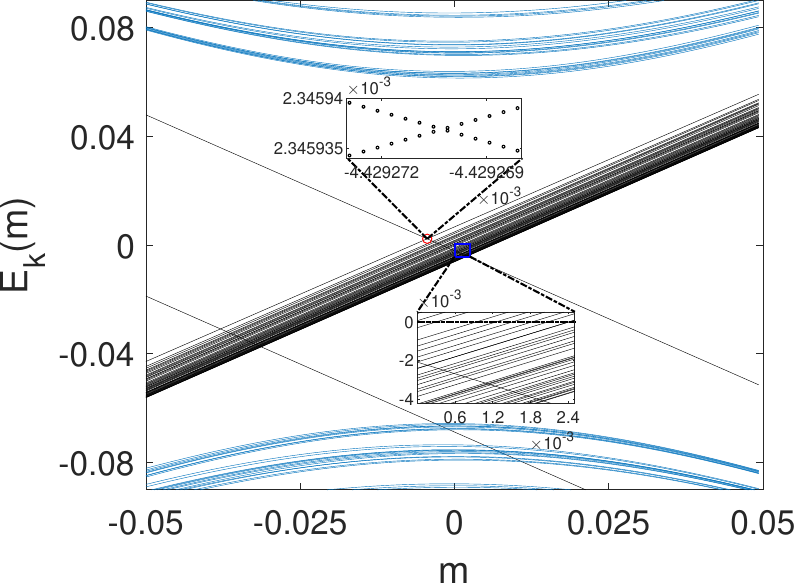}}
	\subfigure[]{\includegraphics[width=0.49\textwidth]{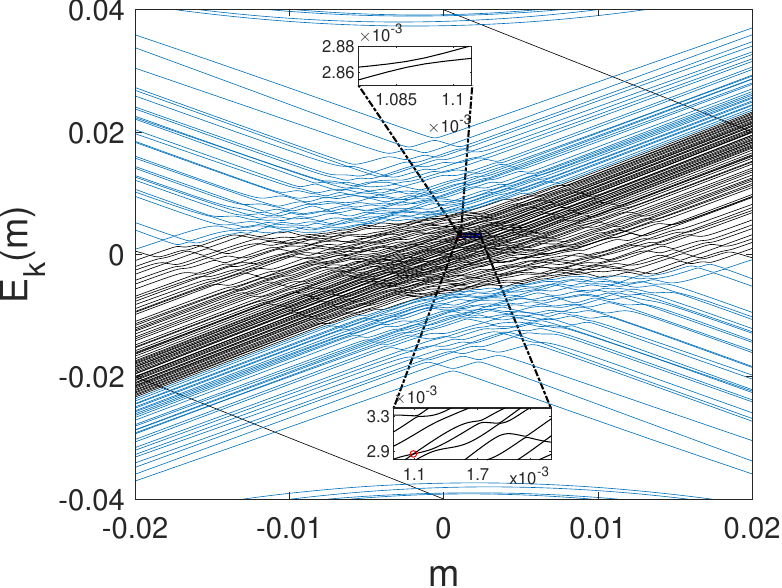}}
	\caption{Spectral flow lines of the eigenvalues $E_k(m)$ of $Q \widetilde{{\cal L}}$, where $k$ labels the different eigenvalues.The blue and black curves denote spectral flow lines $E_k(m)$ with zero and $\pm1$ contributions to the topological index, respectively. (a) Class AIII$_\nu$: $N=14$, $q=4$, $r=2$, $\lambda=0.1$, and $S_+=S_-=-1$. We have $W=\nu=64$ straight lines with positive slope, which, going from $-\infty$ to $+\infty$, contribute to the topological index. As a result, the index $\nu$ equals the number of topological modes. (b) Class CI$_{--\nu}$: $N=14$, $q=4$, $r=1$, $\lambda = 0.01$, and $S=1$. Same features as in (a), but $W = \nu = 128$ flow lines going from $-\infty$ to $+\infty$ contribute to the index.
    (c) Class BDI$^\dag_\nu$: $N=12$, $q=4$, $r=2$, $\lambda = 0.1$, and $S_+=S_-=-1$.
    There exist $48$ flow lines that contribute $+1$ to the topological index, and 16 spectral flow lines (one that is 15-fold degenerate) that contribute $-1$ each, resulting in $W = \nu = 32$. As shown in the upper inset, marked by the red circle, the flow lines with a negative contribution to $\nu$, which are analytical many-body scars~\cite{turner2018}, have true crossings with other flow lines.
	(d) Class BDI$_{++\nu}$: $N=12$, $q=4$, $r=1$, $\lambda=0.02$, and $S=1$. Despite a larger than expected number of real modes, as shown in Fig.~\ref{fig:nrealbdi} (left), the topological index is still the expected one, $W  = \nu = 64$, with 68 and 4 flow lines contributing $+1$ and $-1$ to the topological index, respectively. The four negative-sloped flow lines, which are shown in the plot (in black), are many-body scars that truly cross other flow lines. The red circle in the lower inset marks the region of the upper inset showing that the apparent crossing is an avoided crossing. This illustrates that crossings are avoided unless the state has special properties as, for example, in the case of many-body scars. In the left figures we observe a ``semigap'', which is induced by level repulsion from the zero modes and is of order $\nu/2$ (generally much larger than $N$). For chiral RMT with index $\nu$, it can be shown rigorously that the gap is $\nu/2$~\cite{Shifrin:2005cy}. A similar gap is found in Figs.~\ref{fig:flow2}(d) and \ref{fig:spinchain}. 
    }
        \label{fig:flow1}
        \end{figure*}

\section{Numerical calculation of the topological index}
\label{sec:numerical}

We now confirm the analytical results of the previous section by an explicit calculation of the topological index from the spectral flow.
In Fig.~\ref{fig:flow1}, we depict the $m$ dependence of the eigenvalues $E_k(m)$, referred to as the spectral flow, when $\lambda$ is small, for the four universality classes with expected topological features, see Table~\ref{tab:topology_class}. Black (blue) curves contribute (do not contribute) to the topological index. With $m$ going from $-\infty$ to $+\infty$, a line contributes $+1$ to the topological index $W$ if the corresponding eigenvalue $E_k(m)$ runs from $-\infty$ to $+\infty$, and $-1$ if $E_k(m)$ goes from $+\infty$ to $-\infty$.
In all cases, we have found agreement with the theoretical value of $W$, Eq.~(\ref{eq:topind}), which confirms the topological nature of the dynamics of our model. The spectral flow has a quite rich structure (especially for BDI$^\dagger_\nu$ and BDI$_{++\nu}$), that deserves further analysis.

\begin{figure}[t!]
	\centering
	\includegraphics[width=4.25cm]{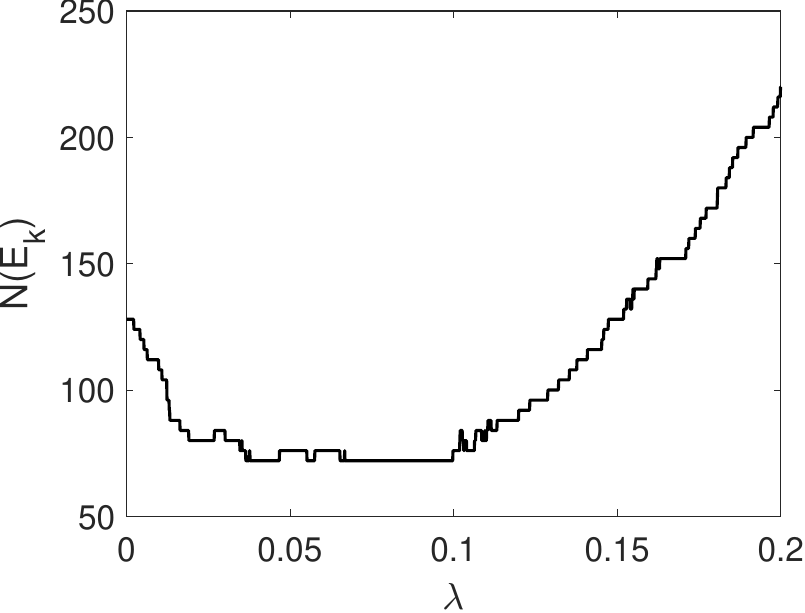}
	\includegraphics[width=4.25cm]{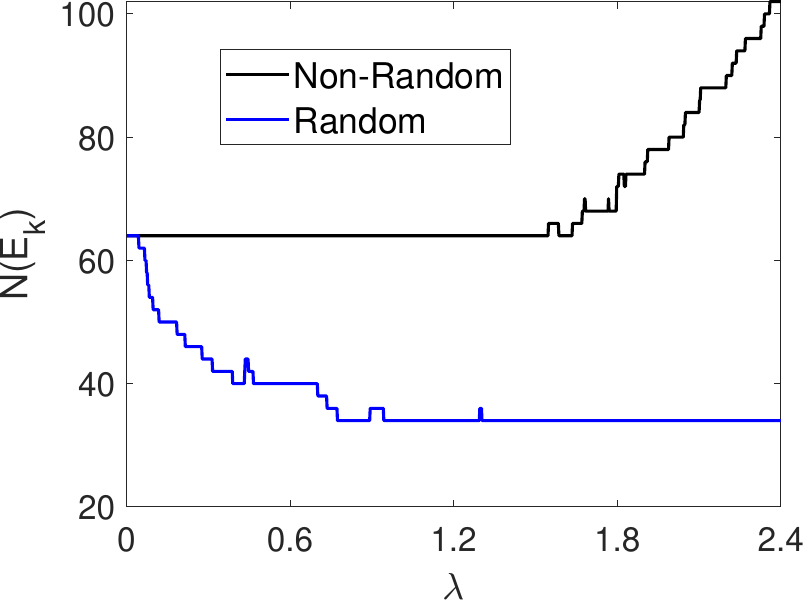}
	\caption{
		Number of exactly real modes as a function of $\lambda$ for a single disorder realization for class BDI$_{++\nu}$ and BDI$^\dag_\nu$.
		(Left) Class BDI$_{++\nu}$: $N=12$, $q = 4$, $r=1$, and $S_+=1$. As $\lambda$ increases from zero, the interaction $H_I$ explicitly breaks the Kramers degeneracy of $H_0$, causing the number of real modes to decrease; when $\lambda$ is between 0.04 and 0.1, only 72 real modes remain, consisting of 68 modes with a positive slope and 4 scar-like states with a negative slope, as illustrated in Fig.~\ref{fig:flow1}(d) and its caption, corresponding to a topological index $\nu=64$.
		(Right) Class BDI$^\dag_\nu$: $N=12$, $q=4$, $r=2$, and $S_+=S_-=1$.
		Black: Contrary to the {\it naive} theoretical prediction that only half of the $64$ real eigenvalues are topological, so that they should not all stay on the real axis as $\lambda$ increases, the number of real modes is independent of $\lambda < 2$.
		Blue: The same but considering a modified $H_I$ Eq.~(\ref{eq:hi}) with Gaussian random couplings
		of zero mean and unit standard deviation. We observe the anticipated sharp decrease for small $\lambda$.
The value of the plateau at intermediate $\lambda$ is equal to 34, comprising 33 with positive flow and one with negative flow, as shown in Fig.~\ref{fig:bdid}, corresponding to a topological index $\nu=32$.}.
	\label{fig:nrealbdi}
\end{figure}

For classes AIII$_\nu$ and CI$_{--\nu}$, see Figs.~\ref{fig:flow1}(a) and \ref{fig:flow1}(b), the lines contributing to $W$ are easily distinguished because they
are also close to straight and run from $-\infty$ to $+\infty$
as a function of $m$, and each contributes $+1$ to $W$.
By contrast, the pattern of the spectral flow for BDI$^\dagger_\nu$ at $N=12$, Fig.~\ref{fig:flow1}(c), and BDI$_{++\nu}$ at $N=12$, Fig.~\ref{fig:flow1}(d), is different, in part due to
Kramers degeneracy of the spectrum for $\lambda = 0$.

\begin{figure}[t]
	\centering
	\includegraphics[width=0.48\textwidth]{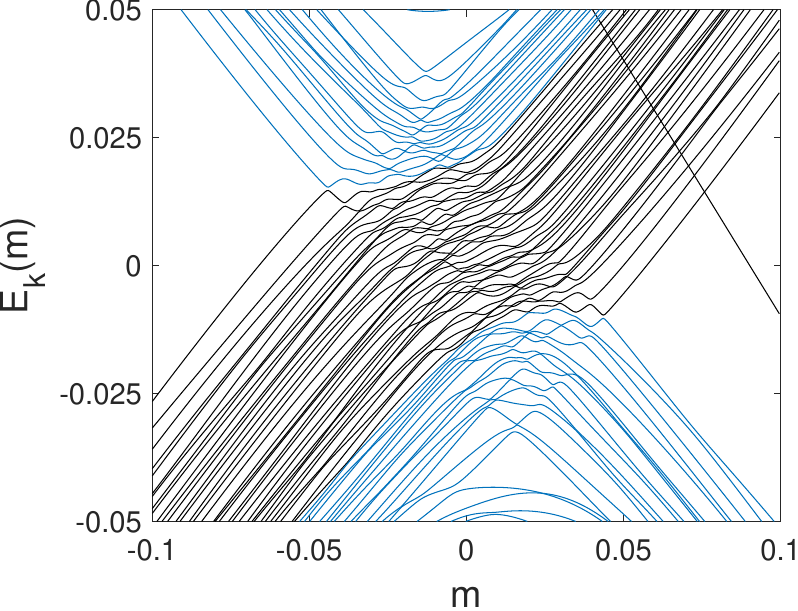}
	\caption{Spectral flow of $E_k(m)$, Eq.~(\ref{eq:ekm}), for class BDI$^\dag_\nu$ ($N=12$, $q=4$, $r=2$, $\lambda = 1$, and $S_+=S_-=1$), after introducing Gaussian random couplings with zero mean and unit standard deviation in $H_I$, Eq.~(\ref{eq:hi}). We find 33 lines with a positive contribution to $\nu$ and one line with a negative contribution which experiences an exact crossing with other flow lines, so that $W = 32 = \nu$.}
	\label{fig:bdid}
\end{figure}

\begin{figure*}[t]
	\centering
	\includegraphics[width=0.49\textwidth]{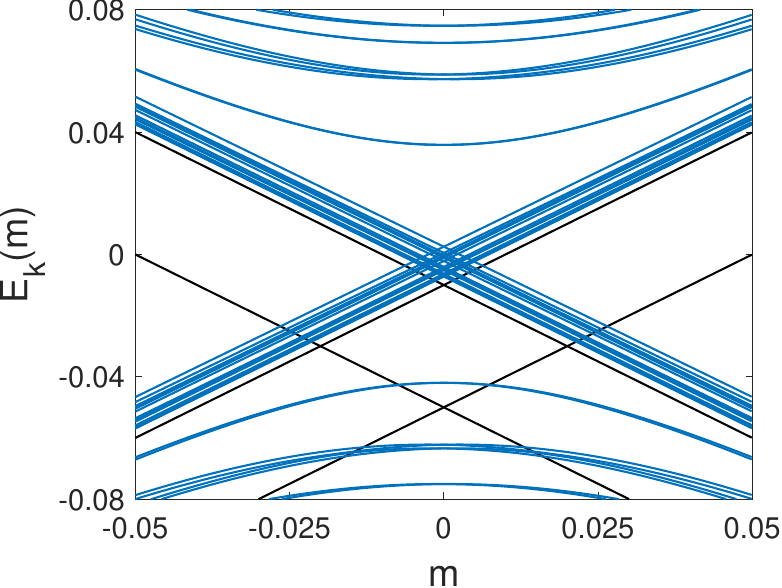}
	\includegraphics[width=0.49\textwidth]{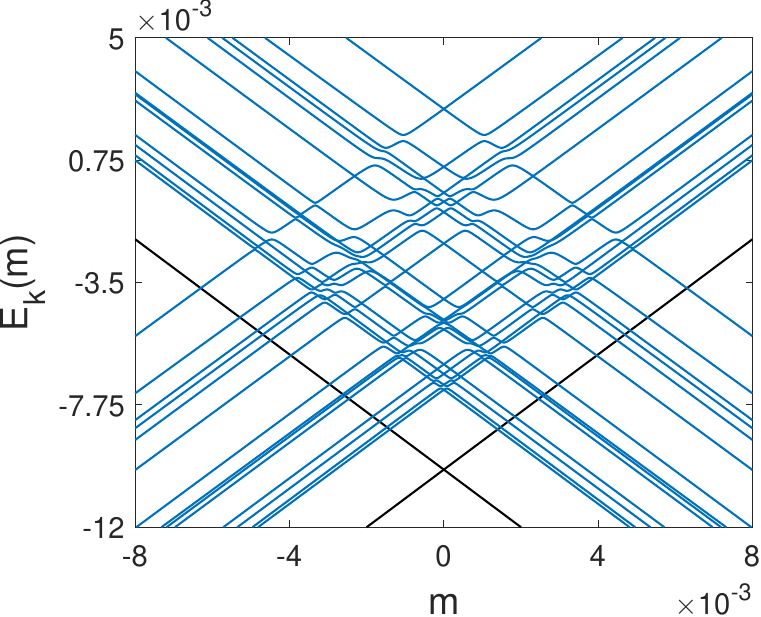}
	\caption{Spectral flow of $E_k(m)$, Eq.~(\ref{eq:ekm}), for class BDI$_{-+}$, with
		$N=10$, $q=4$, $r=1$, $\lambda = 0.01$, $S=-1$, and $m \in [-0.05,0.05]$ (left) or $m \in [-0.008,0.008]$ (right). The topological index is $W = 0$. Although superficially it seems that {\it straight} flow lines in the left panel contribute to $W$, Eq.~(\ref{eq:topind}), at a finer scale (right), many of them show avoided crossings. Moreover, $E_k(m)$ is exactly symmetric around $E_k(0)$ so the topological index is zero.}
	\label{fig:flowS-1}
\end{figure*}

For class BDI$^\dagger_\nu$ and $N = 12$, due to this double degeneracy, we expect $64$ real modes for not too large $\lambda$, $48$ corresponding to flow lines from $-\infty $ to $+\infty$ and 16 corresponding to flow lines from
 $+\infty$ to $-\infty$.
 Therefore, by increasing $\lambda$, we expect the initial $64$ real modes will reduce to only $32$, until their number starts increasing again at a much larger value of $\lambda$ due to additional crossings of the flow lines with the $m$ axis.
 Indeed, see Fig.~\ref{fig:flow1}(c), $W=\Tr \mathbb PQ = \nu = 32$, but the form of the spectral flow is not exactly the expected one. One of the two
 lines in the spectral flow of Fig.~\ref{fig:flow1}(c), going from $+\infty$ to $-\infty$,
 has a fifteen fold degeneracy. Since there are 48 flow lines from $-\infty$
 to $+\infty$, the topological index $W = 48-16 = 32$ agrees with the theoretical expectation.
 Intriguingly, as depicted in the right panel of Fig.~\ref{fig:nrealbdi}, the number of real eigenvalues stays the same as $\lambda$ increases. Only for large values of $\lambda \gtrsim 1$,
 do we observe the expected increase of the number of real eigenvalues, not related to topology, but due to additional
 intersections of the flow lines.
 The reason for this unexpected behavior is that all 16 flow
 lines with a negative contribution to the topological index are related to states that can be obtained analytically or semianalytically, reminiscent of many-body scars~\cite{turner2018}.
 As a consequence of a symmetry particular of $H_I$, these eigenvalues stay on the real axis as $\lambda$ increases.
 Indeed,
 by considering the same $H_I$ but with random couplings in Eq.~(\ref{eq:hi}), which does not alter the class of the full Liouvillian, we observe the anticipated quick decrease of the number of real modes as the coupling to the bath increases, see the right panel of Fig.~\ref{fig:nrealbdi}. Moreover, as illustrated in Fig.~\ref{fig:bdid}, the spectral flow has the $33$ lines from $-\infty$ to $+\infty$ that contribute positively, and a single flow line with no degeneracy from $+\infty$ to $-\infty$ that contributes negatively, resulting in the expected index $W=\nu=32$. In every realization, a single negative flow is linear and truly crossing other lines, which also appears to correspond to a many-body scar, while its position varies due to randomness of different realizations.

\begin{figure*}[t]
	\centering
	\includegraphics[width=0.49\textwidth]{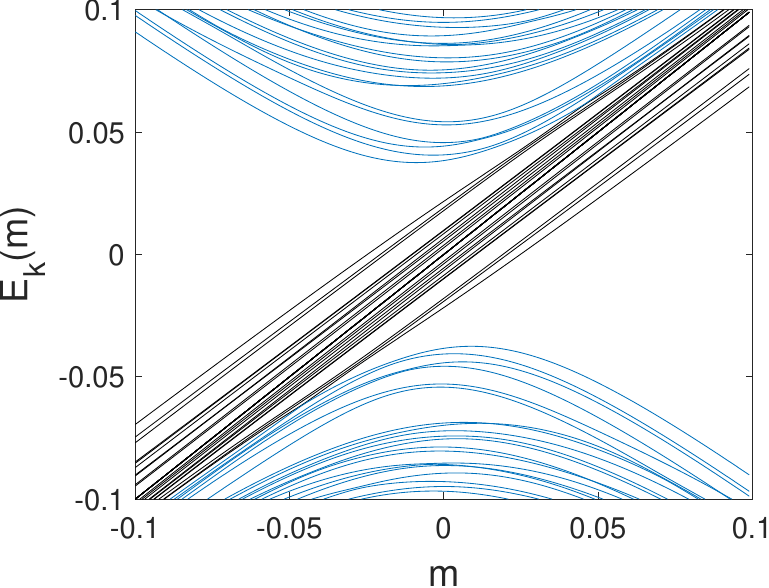}
	\includegraphics[width=0.49\textwidth]{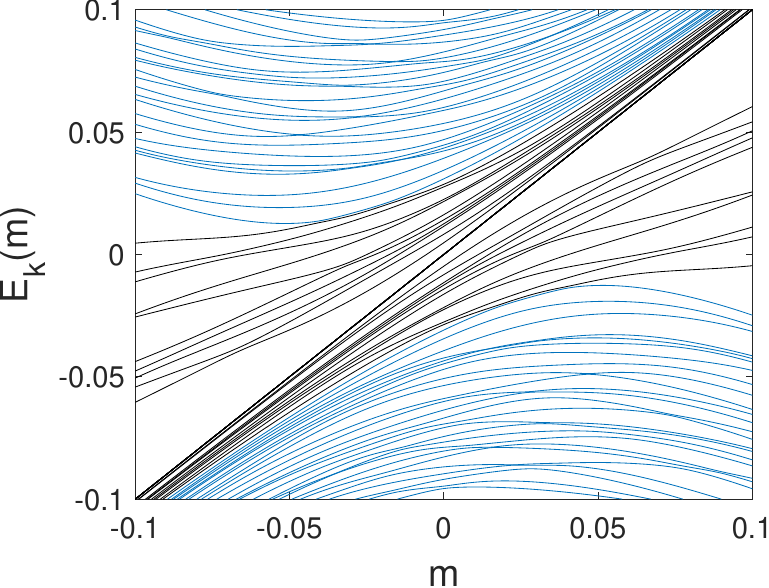}
	\caption{Spectral flow of $E_k(m)$, Eq.~(\ref{eq:ekm}), for class CI$_{--\nu}$, with
		$N=10$, $q=4$, $r=1$, $S=1$, and $\lambda = 0.02$ (left) or $\lambda=0.08$ (right).
		Despite the much more intricate pattern for larger $\lambda = 0.08$, $W = 32 = \nu$ [Eq.~(\ref{eq:topind})] in both cases.
	}\label{fig:flow2}
\end{figure*}

For class BDI$_{++\nu}$, we observe the presence of many almost-straight flow lines from $-\infty$ to $+\infty$ but also from $+\infty$ to $-\infty$
contributing with $+1$ and $-1$, respectively, leading ultimately to the expected topological index $W = 64 = \nu $.
A substantial part of these flow lines does not contribute at all to the topological index
because, after using a much finer resolution, we find that most of the crossings are in fact avoided crossings, see the insets of Fig.~\ref{fig:flow1}(d). Thus, each of these spectral lines flows from $+\infty$ to $+\infty$ or from $-\infty $ to $-\infty$, and
do not contribute to the topological index.
As mentioned before, the origin of the two types of flow lines
is the Kramers degeneracy at
$\lambda = 0$. As a result, the number of zero modes at $\lambda=0$ is $128$
with 96 flow lines from $-\infty $ to $+\infty$ and 32 flow lines from
$+\infty$ to $-\infty$ resulting in an index of $\nu = 64$.
For $\lambda$ very small, most of the zero modes will stay real. However, only the
difference of the two kinds of flow lines
is topologically protected so, as $\lambda$ increases, more real eigenvalues become complex until a minimum of $72$ real eigenvalues, $68$ ($4$) with a positive (negative) contribution to the topological index, is reached, see the left panel of Fig.~\ref{fig:nrealbdi}. Normally, the modes with negative slope would not be protected. In this case, they are not protected by topology but by the fact that they are many-body scars.

Finally, we provide evidence that topology is restricted to
classes for which $\Tr \P_S^s Q$ is nonzero, which only happens for $s=+1$.
For that purpose, we consider the cases where, for each realization, two symmetry classes, indexed by $S=\pm 1$, coexist in the same system. An example is $N/2$ odd and $q/2$ and $r$ even. The $S=1$ block has CI$_{--\nu}$
symmetry with
$\nu = 2^{{N}/{2}}$ while the $S=-1$ block has BDI$_{-+}$ symmetry with $\nu = 0$. The spectral flow in the $S=+1$ sector was obtained in Fig.~\ref{fig:flow1}(d). In the $S=-1$ block, $\P_{S}^{-1}Q \widetilde{{\cal L}}$ has purely imaginary eigenvalues, so we consider instead the spectral flow of $i\P_{S}^{-1}Q \widetilde{{\cal L}}$, for $N=10$, $q=4$, $r=2$, and $S=-1$, see Fig.~\ref{fig:flowS-1}. It is fully symmetric with respect to the $y$ axis. The number of flow lines from $-\infty$ to $ +\infty$ is equal to the number
of flow lines from $+\infty$ to $-\infty$ resulting in a vanishing topological
index. The spectral flow lines are symmetric about $m=0$, as shown in the right panel of Fig.~\ref{fig:flowS-1}, such that the index is actually zero. However, there are mostly avoided crossings, while two analytic modes truly cross other flow lines.

\begin{figure}[t]
	\centering
	\includegraphics[width=0.48\textwidth]{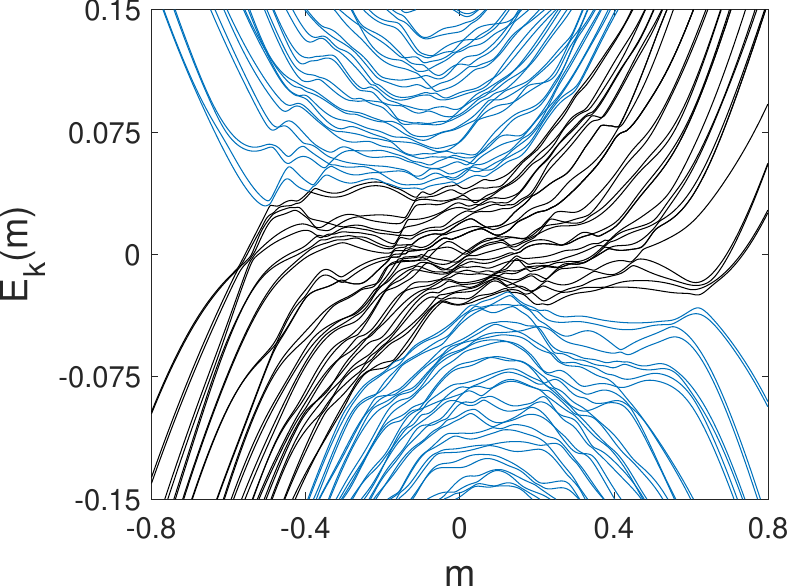}
	\caption{Spectral flow of $E_k(m)$, Eq.~(\ref{eq:ekm}), considering a single-site SYK model Eq.~(\ref{eq:single_syk}) with both real and imaginary random couplings of the same strength ($\kappa=1/2$),
		for $N=10$, $q=4$, $r=1$, $\lambda = 0.01$, and $S=1$. Although the corresponding symmetry class changes from CI$_{--\nu}$ to BDI$^\dag_\nu$, the topological index is still the same, $\nu=W = 32$, as in the case of purely real random couplings considered in the rest of the paper.}\label{fig:k05}
\end{figure}

\section{Universality of topology}
\label{sec:universality}

Having provided convincing evidence of the topological nature of the real modes for certain universality classes of the
SYK model coupled to a bath,
we now turn to the discussion of the robustness and the universality of these results in different realms.

\subsection{Topology and large $\lambda$}

We expect that increasing $\lambda$ does not affect the topological
index as the eigenvalues are continuous functions of $m$ and $\lambda$.
As an example, the spectral flow in Fig.~\ref{fig:flow2} for class CI$_{--}^{\nu}$ is quite different for $\lambda=0.02$ and $0.08$, but, of course, we still obtain the same topological index $W=\nu$. The pattern for
$\lambda = 0.08$ is much more intricate, with flow lines that
may cross the $m$ axis multiple times while almost straight lines --- typical for $\lambda = 0.02$ --- are missing.
Moreover, as can be deducted from the number of crossings
of the flow lines with the $m$ axis, the number of purely real eigenvalues for $\lambda = 0.08$ has increased with respect to $\lambda = 0.02$, which may lead to the suspicion that the topological index has changed. However, this is not the case.
As expected, we still find that $W = \nu$ for {\it each} disorder realization. This is a further confirmation that $\nu$ is a topological invariant independent of the total number of real modes or the coupling $\lambda$.

\subsection{Dynamical generators beyond the Lindblad form}

It is also clear from the symmetry classification in Table~\ref{tab:topology_class} that topology depends only on $r{\rm mod} 2$ and $N{\rm mod}4$,
so perturbations that remain in a given topological class will lead to an
identical topological invariant $W = \nu$. Likewise, the presence of topology is insensitive to the addition of a purely real random part to the imaginary random couplings in Eq.~(\ref{eq:single_syk}),
$K_{i_1\cdots i_q}\to -i\sqrt{1-\kappa} J_{i_1\cdots i_q}+ \sqrt{\kappa} K_{i_1\cdots i_q}$, with $\kappa\in[0,1]$ parametrizing the degree of non-Hermiticity of the couplings~\cite{garcia2023c},
provided that the pseudo-Hermiticity of $\cal L$ in Eq.~(\ref{eq:Keldysh_twosite SYK}) is preserved (although the non-Hermitian symmetry class could change). Results for the spectral flow for $\kappa = 1/2$, depicted in Fig.~\ref{fig:k05}, confirm the robustness of topology in this case, as the topological invariant $W=\nu$ remains unchanged.
We note this addition of random real couplings corresponds to considering dissipative dynamics that, contrary to Lindbladian dynamics, do not preserve the
trace of the density matrix. In Ref.~\cite{garcia2023c}, we recently showed that such dynamical generators can be mapped to PT-symmetric Hamiltonians and, therefore, also those non-Hermitian systems display topological behavior.

\begin{figure*}[t]
	\centering
	\includegraphics[width=\textwidth]{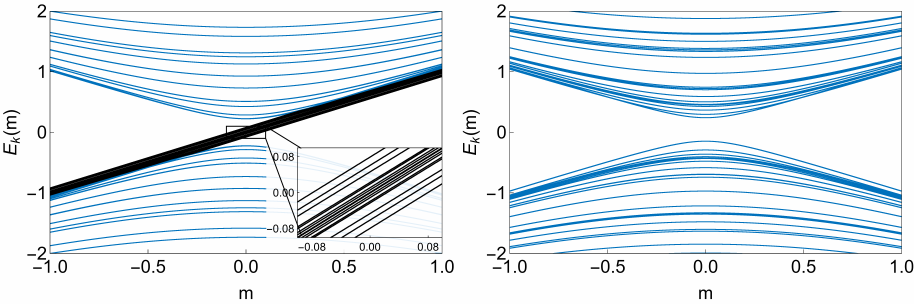}
	\caption{Spectral flow of $E_k(m)$, Eq.~(\ref{eq:ekm}), for the dephasing spin chain of Eqs.~(\ref{eq:spin_H})--(\ref{eq:spin_calL}) with $L=4$ and dissipation strength $\gamma$ from a box distribution in $[0.005, 0.015]$. Left: Even parity sector, corresponding to class BDI$_{++\nu}$. The pseudo Hermiticity operator \textsc{swap} develops an anomalous trace ($\nu=16$) which matches the topological index $W$, namely, there are exactly $W=\nu=16$ lines flowing from $-\infty$ to $+\infty$. Right: Odd parity sector, corresponding to class CI$_{+-}$. \textsc{swap} is not anomalous and there are no signatures of topology in the spectral flow.}
	\label{fig:spinchain}
\end{figure*}

\subsection{Non-SYK Lindbladians}

We now show that topology is not particular to the SYK setting by computing the spectral flow associated
with $\cal L$ for a spin chain in the same symmetry class, see Fig.~\ref{fig:spinchain}.
More precisely, we consider a spin-$1/2$ chain on a 1D lattice with $L$ sites, which we represent by Pauli matrices $\sigma^\alpha_j$, $\alpha=x,y,z$, $j=1,\dots,L$. Specifically, we choose a Liouvillian ${\cal L}$ that couples a nearest-neighbor XYZ Heisenberg chain in a transverse field,
\begin{equation}
	\label{eq:spin_H}
	H=\sum_{j=1}^{L-1} \left( J_j^x \sigma^x_j \sigma^x_{j+1}
	+ J_j^y \sigma^y_j \sigma^y_{j+1}
	+ J_j^z \sigma^z_j \sigma^z_{j+1}\right)
	+\sum_{j=1}^L g_j \sigma^x_j
\end{equation}
to an environment characterized by dephasing jump operators,
\begin{equation}
	L_j = \sqrt{\gamma_j} \sigma^z_j.
\end{equation}
For the numerical results, we set $L = 4$, $J^x_j = 1$, $J^y_j=0.8$, $J^z_j=0.55$, and sample $g_j$ and $\gamma_j$ uniformly from box distributions in $[-0.7,0.7]$ and $[0.005,0.015]$, respectively. Note that this places the chain in the weakly-dissipative regime (equivalent to small $\lambda$ in the SYK setting).

The Lindbladian is vectorized as
\begin{equation}
	\label{eq:spin_calL}
	\sL = -iH \otimes \id + \id \otimes iH + \sum_{j=1}^L \gamma_j \sigma^z_j \otimes \sigma^z_j,
\end{equation}
where we have again dropped an irrelevant constant. The pseudo-Hermiticity operator, $Q\sL^\dagger=\sL Q$ is $Q=\textsc{swap}$, where \textsc{swap} is defined as $\textsc{swap} (A\otimes B) \textsc{swap}=B\otimes A$ for any operators $A$ and $B$. In terms of Pauli matrices, it can be written as
\begin{equation}
	Q=\prod_{j=1}^L \frac{1}{2}(\id_j\otimes \id_j+\sigma_j^x \otimes \sigma_j^x+\sigma^y_j\otimes \sigma^y_j+\sigma^z_j\otimes \sigma^z_j).
\end{equation}
The Lindbladian has $\mathcal{U}=\prod_{j=1}^L \sigma^x\otimes \sigma^x$ as a
symmetry ($[\sL,\mathcal{U}]=0$) and the block diagonalizes into two sectors labeled by the parity $\mathcal{U}=\pm1$. As before, we must fix one of these sectors and project $\sL$ and $Q$ into it.
In the even parity sector, the spin chain belongs to class BDI$_{++\nu}$ (an exhaustive symmetry classification of such spin chains was provided in Ref.~\cite{sa2022a} but their topological nature was not identified), \textsc{swap} projected into this sector develops an anomalous trace $\nu=2^L$ and,
computing the spectral flow, we find, as expected, the topological index $W = \nu = 2^L$. On the other hand, in the odd parity sector, the Lindbladian belongs to the non-topological class CI$_{+-}$, because the \textsc{swap} operator projected into this sector has vanishing trace. These findings are confirmed by the computation of the spectral flow plots depicted in Fig.~\ref{fig:spinchain}.

\begin{figure*}[t]
	\centering
	\includegraphics[width=0.49\textwidth]{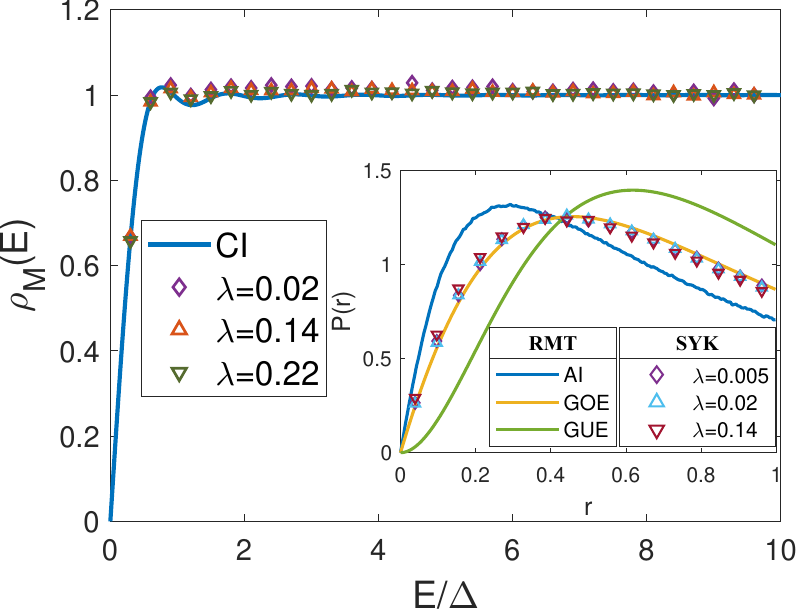}
	\includegraphics[width=0.49\textwidth]{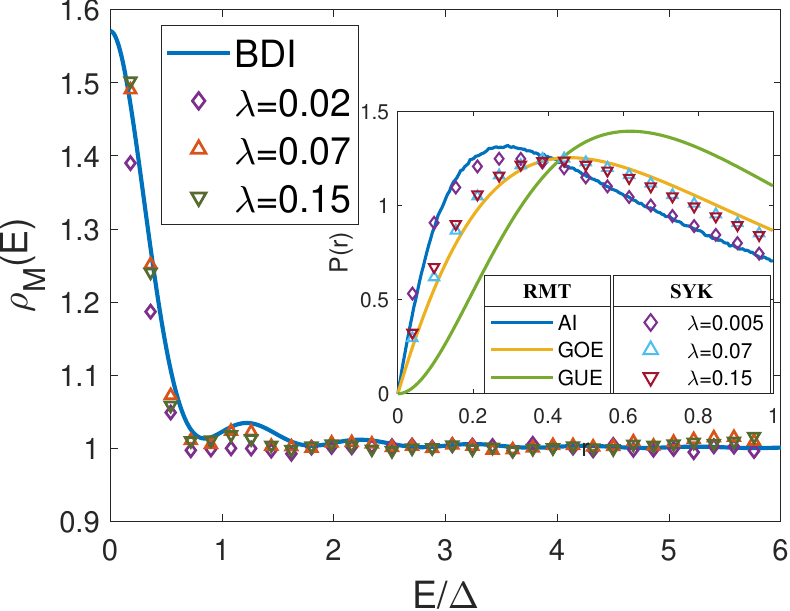}
	\caption{
		(Left) Level statistics for the real modes of $\cal L$ Eq.~(\ref{eq:Keldysh_twosite SYK}) for class CI$_{--\nu}$ with $N=14$, $q=4$, $r=1$, and $S=1$. (Main) Microscopic spectral density $\rho_M(E)$ of the real modes close to $E=0$ in units of the mean level spacing $\Delta$ from $2.5 \times 10^5$ realizations. (Inset) Distribution of the gap ratio $P(r)$ with $10^4$ realizations in the bulk of the spectrum. We observe excellent agreement with the CI random matrix results for all $\lambda$.
		(Right) Level statistics of the real modes of ${\cal L}$ Eq.~(\ref{eq:Keldysh_twosite SYK}) for class BDI$_{++\nu}$ with $N=12$, $q=4$, $r=1$, and $S=1$. (Main) The microscopic spectral density $\rho_M(E)$ (obtained from $10^6$ realizations) shows good agreement with the BDI random matrix prediction for $\lambda$ corresponding to the region where most real eigenvalues are topological.
		(Inset) Distribution of the gap ratio $P(r)$ with $4\times 10^4$ realizations in the bulk of the spectrum which shows good agreement with RMT in the expected region of parameters.
		 $\lambda=0.07$ corresponds to the plateau region in Fig.~\ref{fig:nrealbdi} (left), where there are 72 real modes, comprising 68 with positive flow and 4 with negative flow.}
\label{fig:Pr_normal}
\end{figure*}

\begin{figure*}[t]
\centering
\includegraphics[width=0.49\textwidth]{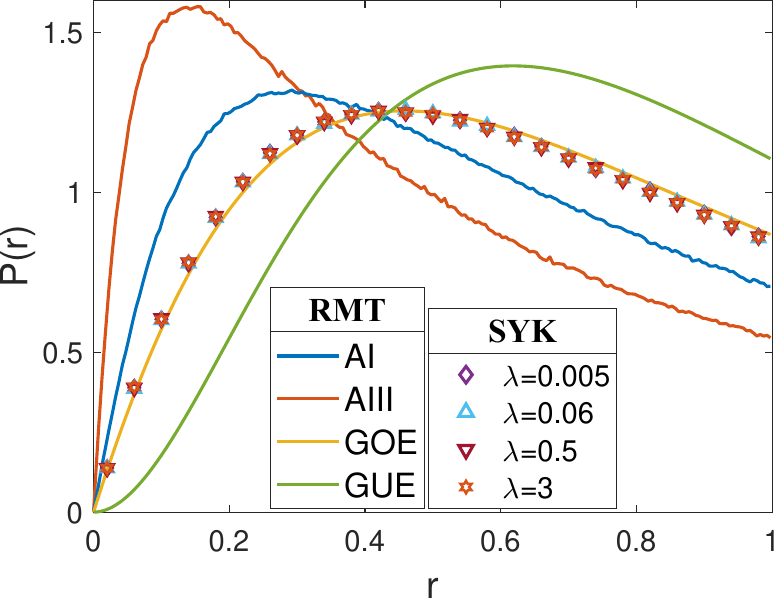}
\includegraphics[width=0.49\textwidth]{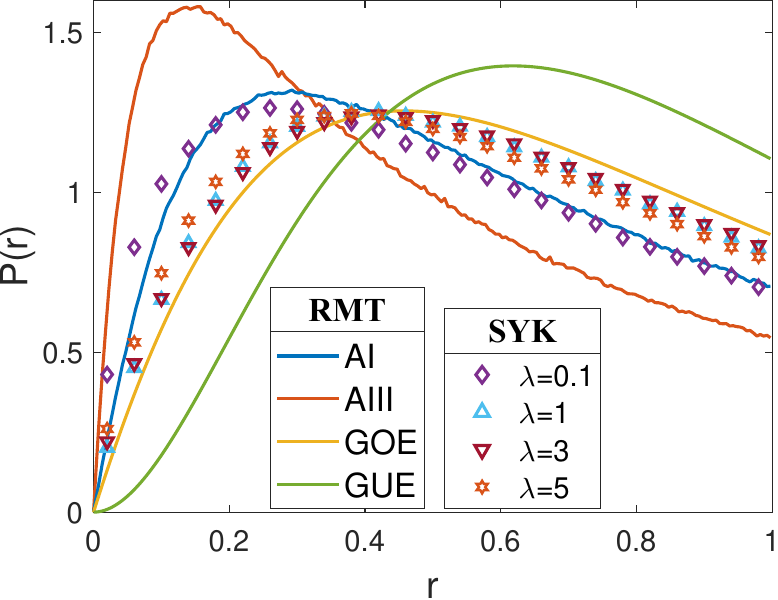}
\caption{
	(Left) Gap ratio statistics $P(r)$ for the bulk real eigenmodes of $\cal L$ Eq.~(\ref{eq:Keldysh_twosite SYK}) with $N = 12$, $q = 4$, $r = 2$, and $8 \times 10^4$ realizations, corresponding to class BDI$^\dagger_\nu$. We find excellent agreement with the BDI random matrix prediction as most real eigenvalues are topological. We note that the 16 eigenvalues with a negative contribution to $\nu$ are related to many-body scars, and therefore are excluded from the level statistics analysis.
	(Right) Same parameters as in the left plot but using random couplings in $H_I$, Eq.~(\ref{eq:hi}). For $\lambda=1,3$, which correspond to the plateau region of Fig.~\ref{fig:nrealbdi} (right), we select $8\times 10^4$ realizations that exhibit the minimum count of 34 real modes, composed of 33 with positive flow and one with negative flow.
	When the value of $\lambda$ is such that the number of real modes approaches $\nu$, the agreement with the GOE prediction improves. The remaining deviations for intermediate values of $\lambda$ may be due to the effect of two eigenvalues related to many-body scars that are included in the level statistics analysis.}
\label{fig:Pr_abnormal}
\end{figure*}

Having worked out two explicit examples, we can put forward a general sufficient condition for the emergence of topology in open many-body quantum systems described by a general Hamiltonian $H$ and a set of jump operators $L_\mu$. Let us consider the vectorization used for the spin chain (similar considerations apply for the fermionic vectorization). Before vectorization, $Q$ is a superoperator that implements transposition, $Q(\rho)=\rho^T$ for any operator $\rho$. Composing it twice with $\sL$ yields, after some straightforward algebra:
\begin{equation}
\begin{split}
	Q\sL Q(&\rho)=+i[H^*,\rho]+
	\sum_{\mu}\left[ L^*_\mu \rho L_\mu^T -\frac{1}{2}\{(L_\mu^\dagger L_\mu)^*,\rho\}\right].
\end{split}
\end{equation}
Comparing it with the action of the adjoint Lindbladian,
\begin{equation}
	\sL^\dagger(\rho)=+i[H,\rho]
	+\sum_{\mu}\left( L^\dagger_\mu \rho L_\mu -\frac{1}{2}\{L_\mu^\dagger L_\mu,\rho\}\right),
\end{equation}
and using $Q^2=1$, we find that we obtain the desired pseudo-Hermiticity symmetry $Q \sL^\dagger=\sL Q$ if the Hamiltonian is real and all jump operators are Hermitian and real up to a phase. (The condition on the jump operators can be relaxed to a normality condition and an appropriate rotation of $Q$~\cite{sa2022a}, but Hermitian jump operators are a standard choice of dissipation and enough for our purposes.)
We have thus shown that the mechanism for the emergence of topology is the existence of a pseudo-Hermiticity operator with anomalous trace. In Lindbladian systems, the operator $Q$ exchanging the bra and ket spaces of the density matrix (or Keldysh contours) is the most natural candidate, as we demonstrated explicitly.

\subsection{Dynamical and experimental signatures}

A major appeal of topology is its experimental relevance. For instance, in systems experiencing the integer or spin Hall effect, linear response transport is governed by a topological invariant.
By contrast, topology of the vectorized Liouvillian reveals itself in the full out-of-equilibrium time evolution.
Indeed, the topological real modes have a sharp dynamical interpretation.
These modes control the approach to a steady state of the system independently of the coupling to the bath $\lambda$ provided that it is weak enough so that all real modes are topological.
A promising computational scheme to reveal these topological features in the dynamics is that of subspace-Krylov techniques that may allow reach a sufficiently large value of $N$
to compare with the non-topological saddle-point results for the decay rate~\cite{garcia2022e}. 

Regarding experimental signatures of many-body topology, we note that 
the studied topological real modes have positive parity and will couple to operators representing observables of the same parity. An example of an observable with these properties is the susceptibility described by the commutator $\theta(t) [S_x(0),S_x(t)]$ with $S_x$ a spin operator. In the long-time limit, the susceptibility is closely related to the diffusion coefficient, which will certainly be influenced by the topological real modes that control that dynamical region. More generally, the decay rate to the equilibrium value of any operator with even parity will be totally or mostly controlled by topology for sufficiently weak dissipation. A direct experimental measure of the decay rate is feasible by using quantum optical settings~\cite{schomerus2013} with the appropriate symmetry. Indeed, slowly relaxing many-body states and subdiffusive dynamics have already been observed experimentally~\cite{bouganne2020} in a gas of strongly interacting quantum bosons with dissipative effects induced by spontaneous emission.

\subsection{Topology and level statistics of real modes}

We now turn to another playground to probe topology: level statistics of the real eigenvalues.
As discussed above, there are two sources of real eigenvalues, topological (which are robust to changes of $\lambda$) and nontopological (which are sensitive to $\lambda$). These two types of eigenvalues do not necessarily have the same spectral statistics.
We have recently found~\cite{Akemann:2010em,Kieburg:2015vqa,garcia2023c} that, perturbatively ($\lambda \ll 1$), level statistics of the topological real modes for classes AIII$_\nu$, BDI$^\dagger_\nu$, CI$_{--\nu}$, and BDI$_{++\nu}$ are given by those of random Hermitian matrices belonging to the Gaussian unitary ensemble (GUE), Gaussian orthogonal ensemble (GOE), CI, and BDI universality classes, respectively. On the other hand, the statistics of nontopological real eigenvalues follow the predictions of the respective non-Hermitian universality classes (AIII, BDI$^\dagger$, CI$_{--}$, and BDI$_{++}$, respectively).
In this section, we test the robustness of these results in our SYK setting.

We first consider CI$_{--\nu}$ at $N = 14$, corresponding to $q = 4, \;r = 1$ where,
for small $\lambda$, the number of exact real modes is equal to the number of topological modes. In that case, we have found, both in the bulk --- by the calculation of the distribution of the gap ratio $P(r)$~\cite{atas2016} --- and close to the origin --- where we compute the microscopic spectral density $\rho_M(E)$ --- an excellent agreement with the RMT prediction for class CI, see Fig.~\ref{fig:Pr_normal} (left). Also for class AIII$_\nu$, all real eigenvalues are topological for not very large $\lambda$ and we find (not shown) a very similar level of agreement with the predicted GUE class~\cite{garcia2023c}.

As mentioned earlier, the situation is more complicated for BDI$_{++\nu}$. Due to the Kramers degeneracy at $\lambda = 0$, for sufficiently small $\lambda$, the number of topological modes is only about half the total number of real modes, see the left plot of Fig.~\ref{fig:nrealbdi}.
Therefore, we expect deviations from the RMT prediction for small $\lambda$ and find agreement once the number of real eigenvalues is equal to $\nu=\Tr \mathbb{P} Q$.
The results depicted in Fig.~\ref{fig:Pr_normal} (right) largely confirm this picture. In the bulk of the spectrum, $P(r)$ deviates strongly from the BDI prediction (i.e., GOE statistics) for very small $\lambda$. The agreement with BDI gradually improves as $\lambda$ increases and most real eigenvalues are topological.
Similar level of agreement occurs for the microscopic spectral density $\rho_M(E)$, namely, there is good agreement with the random matrix result in class BDI, provided that the number of real modes is close to $\nu$.

The remaining class, BDI$^\dagger_\nu$, also has Kramers degeneracy, so superficially one would expect similar deviations as in class BDI$_{++\nu}$.
However, $P(r)$, depicted in the left panel of Fig.~\ref{fig:Pr_abnormal}, shows excellent agreement with the predicted GOE result for a broad range of values of $\lambda$. Moreover, unlike
the BDI$_{++\nu}$ class, the number of real modes, see the left panel of Fig.~\ref{fig:nrealbdi}, does not depend on $\lambda$ unless it is very large ($\lambda > 2$) and the direct effect of the bath pushes eigenvalues onto the real axis.
We removed two real modes, one with a fifteen fold degeneracy, from the level statistics calculation because they seem to be related to many-body scars~\cite{turner2018} and, therefore, are not quantum chaotic. The degenerate flow lines with negative slope have exact crossings with the flow lines with a positive slope and cannot be deformed into pairs of flow lines that do not intersect the $m$ axis. Even though there are thus more real modes than the value of the topological index, we still observed agreement with GOE level statistics.
As a confirmation of this point, we repeat the calculation of $P(r)$ for random couplings in $H_I$, Eq.~(\ref{eq:hi}), so that the symmetry is still BDI$^\dagger_\nu$. In this case, the dependence on the number or real modes with $\lambda$, see the right panel of Fig.~\ref{fig:nrealbdi}, is very similar to that of class BDI$_{++\nu}$. Not surprisingly, the $\lambda$ dependence of $P(r)$, see the right panel of Fig.~\ref{fig:Pr_abnormal}, mimics that of the BDI$_{++\nu}$ class. The slightly larger deviation for intermediate $\lambda$ values may be due to the single mode related to scars that are included in the level statistics analysis.

These results point to a nuanced relation between topology and level statistics
 if not all real modes are topological as in BDI$_{++\nu}$ and BDI$^{\dagger}_\nu$ classes.
 Therefore, a symmetry classification of many-body non-Hermitian topology by level statistics is feasible but likely will require extra effort to identify topology just by a level statistics analysis.

\section{Conclusion and outlook}
\label{sec:conclusion}

For four pseudo-Hermitian symmetry classes, AIII$_\nu$, BDI$^\dagger_\nu$, CI$_{--\nu}$, and BDI$_{++\nu}$~\cite{garcia2023c}, we have found robust topological features
in the dynamics of a single site SYK weakly coupled to a Markovian bath, which is a paradigmatic model of many-body dissipative quantum chaos.
The route to recognizing topology in this setting starts with the observation that the trace of the projection of a unitary operator $Q$, related to the pseudo-Hermiticity of the vectorized Liouvillian, is non-vanishing.
Specifically, after projection onto states with even parity, $Q$ develops an anomalous trace, $\Tr\mathbb{P} Q = \nu \ne 0$. As is the case for the Dirac
operator for gauge fields with a nonzero topological charge, in the limit of zero coupling, the vectorized
Liouvillian admits a representation with diagonal blocks of zeros of unequal size leading to exact zero modes.
For small $\lambda$, we have shown those zero modes become real modes of the Liouvillian. We stress that this rectangularity is an emergent property and is not introduced ad hoc, as was the case in earlier random matrix studies~\cite{Akemann:2010em,Kieburg:2015vqa}.

Using spectral flow, we have shown, analytically and numerically, that for these four classes the index $ \Tr \mathbb{P}Q = \nu$ is a topological invariant that is stable with respect to deformations of the Liouvillian.
The topological structure is also robust to the replacement of the SYK model by a spin chain with the same symmetry or to the addition of a non-Hermitian part to the single-site SYK, which illustrates the universality of our results.
 Physically, the topological invariant can be interpreted as a pseudo-Hermiticity quantum anomaly, represented by the exchange operator $Q$, reminiscent of the chiral anomaly in QCD.

 Topology leaves an imprint on the level statistics of real modes though the degree of universality is limited by the presence, in some cases, of real modes not related to topology.
In the limit of weak coupling, these topological modes control the dynamics for late times. Therefore, anomalies in the decay rate to the steady state are fingerprints of topology.
We stress our findings are unrelated to the physics of topological insulators or
its non-Hermitian analog since in our case interactions are crucial
for the existence of topological states.

Natural extensions of our work include a full classification of topological universality classes, out of the 38~\cite{ueda2019,bernard2002} for non-Hermitian systems, and the effect
of topology on equilibration.
Generalizations to higher dimensions or the use of tensor~\cite{witten2016,gurau2017,krishnan2017,Kim:2019upg}, supersymmetric~\cite{fu2018,kanazawa2017,garcia2018a,li2017} or Wishart~\cite{sa2022b} SYK models are natural choices to further explore topology in many-body dissipative quantum chaos.

\acknowledgments{
Useful discussions with Shu-Heng Shao are acknowledged.
A.\ M.\ G.\ and C.\ Y.\ acknowledge support from the
National Natural Science Foundation of China (NSFC):
Individual Grant No. 12374138, Research Fund for
International Senior Scientists No. 12350710180, and National
Key R$\&$D Program of China (Project ID:
2019YFA0308603). A.\ M.\ G.\ acknowledges support
from a Shanghai talent program. L.\ S.\ was supported by a Research Fellowship from the Royal Commission for the Exhibition of 1851 and by Fundação para a Ciência e a Tecnologia (FCT-Portugal) through Grant No.\ SFRH/BD/147477/2019. J.\ J.\ M.\ V.\ acknowledges support from U.S.\ DOE Grant No. DE-FAG88FR40388. L.\ S.\ and J.\ J.\ M.\ V.\ acknowledge hospitality and support from the Simons Center for Geometry and Physics and the program ``Fluctuations, Entanglements, and Chaos: Exact Results'' and the workshop
``Symmetric Mass Generation, Topological Phases and Lattice Chiral Gauge Theories'',
where some of the ideas of this paper were developed.
}

\bibliography{librarynh_amg.bib}

\end{document}